\documentclass[11pt]{article}
\usepackage{moriond,epsfig}

\def\Journal#1&#2&#3(#4){#1{\bf #2}, #3 (#4)}

\def\NPB{Nucl.  Phys.  {\bf B}}
\def\PLB{Phys.  Lett.  {\bf B}}
\def\PRL{Phys.  Rev.  Lett.  }
\def\PRD{Phys.  Rev.  {\bf D}}

\def\etal{{\it et al.}}

\def\be{\begin{equation}}
\def\ee{\end{equation}}
\def\bea{\begin{eqnarray}}
\def\eea{\end{eqnarray}}

\newcommand{\rar}{\rightarrow}
\newcommand{\jpsi}{J/\psi}
\newcommand{\psip}{\psi(2S)}
\newcommand{\psipto}{\psi(2S)\rightarrow}

\newcommand{\pipipi}{\pi^+\pi^-\pi^0}
\newcommand{\gamg}{\gamma\gamma}

\newcommand{\wpi}{\omega\pi^0}

\newcommand{\eeto}{e^{+}e^{-}\rightarrow}
\newcommand{\kstark}{K^*(892)\bar{K}+c.c.}
\newcommand{\kstarkpm}{K^*(892)^+K^-+c.c.}
\newcommand{\kstarknn}{K^*(892)^0\bar{K}^0+c.c.}
\newcommand{\ks}{K_s^0}
\newcommand{\GeV}{~\hbox{GeV}}

\newcommand{\etap}{\eta^{\prime}}

\newcommand{\rhoto}{\rho(2150)}

\begin{document}
\vspace*{4cm}
\title{RECENT RESULTS ON $\psip$ DECAYS AT BES}

\author{ L. L. MA \\ (for the BES collaboration) }

\address{Institute of High Energy Physics, Chinese Academy of Sciences,\\
Beijing 100049, China}

\maketitle\abstracts{
Recent results on $\psip$ decays, including 10 Vector + Pseudoscalar (VP)
modes and $p\bar{p}\pi^0(\eta)$, are reported with $14\times10^6$ 
$\psip$ events collected with the BESII detector. Cross sections and 
form factors for $e^+e^- \to \wpi$, $\rho\eta$, and $\rho\etap$ at the
center of mass energies of 3.650, 3.686, and 3.773 GeV are measured
 simultaneously.}

\section{Introduction}
A strong violation to the ``12\% rule'' predicted by perturbative QCD
 was first observed by the
MarkII experiment in the Vector-Pseudoscalar (VP) meson  final states,
$\rho\pi$ and $K^{*+}(892)K^{-}+c.c.$~\cite{markII}. Significant suppressions
observed in four Vector-Tensor decay modes~\cite{BESII_VT} make the puzzle even
more mysterious.  Numerous theoretical explanations have been
suggested~\cite{qcd15_th}, but the puzzle still remains one of the most
intriguing questions in charmonium physics.

The study on $\psipto p\bar{p}\pi^0(\eta)$ provides a chance to study
the $N^*$ resonances, which play important roles on our understanding 
of the nonperturbative QCD.

\section{Analysis of $\psipto\pipipi$}
\begin{figure}[htbp]
\centerline{\hbox{\psfig{file=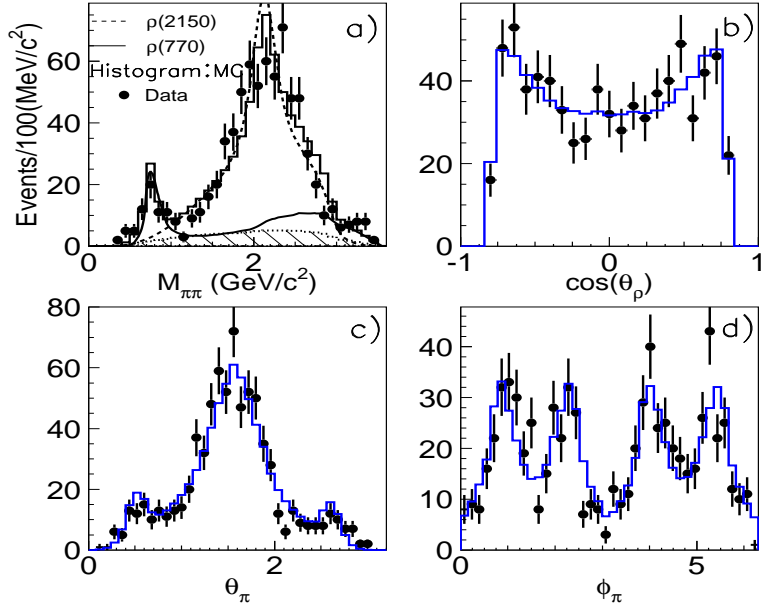,height=8cm, width=10cm}}}
\caption{ Comparison between data (dots with error bars) and the
final fit (solid histograms) for (a) two pion invariant mass, with
a solid line for the $\rho(770)\pi$, a dashed line for the
$\rhoto\pi$,  and a hatched histogram for background; (b) the
$\rho$ polar angle in the $\psip$ rest frame; and (c) and (d) for the polar
and azimuthal angles of the designated $\pi$ in $\rho$ helicity frame. }
\label{fit}
\end{figure}

The selected $\pipipi$ events are fitted in 
the helicity amplitude formalism with an unbinned maximum likelihood 
method using MINUIT~\cite{mini}. The fit shown in Fig. \ref{fit}
 describes the data reasonably 
well, and the $\rhoto$ serves as an effective description of the high
mass enhancement near 2.15~GeV/$c^2$ in $\pi\pi$ mass~\cite{wangz}.
The branching fractions of $\psipto \pi^+ \pi^-
\pi^0$, $\rho(770)\pi$ and $\rho(2150)\pi \rar \pi^+ \pi^- \pi^0$ 
are $(18.1 \pm 1.8 \pm 1.9)\times 10^{-5}$
, $(5.1\pm 0.7\pm 1.1) \times 10^{-5}$ and 
$(19.4 \pm 2.5 ^{+11.5}_{-3.4}) \times10^{-5}$, respectively, 
where the first error is statistical and the second one
is systematic. 

\section{Analysis of Electromagnetic Decays $\psipto\omega\pi, \rho\eta$ and 
$\rho\etap$ }
For this analysis, beside the $\psip$ data sample, we also analyze
6.42 pb$^{-1}$ of continuum data at $\sqrt{s}=3.650\GeV$~\cite{Lcont}, and
17.3 pb$^{-1}$ at the $\psi(3770)$~\cite{Lpsipp}.
Table \ref{formfactor} lists the cross sections of $\eeto\omega\pi, \rho\eta$
 and $\rho\etap$ and the corresponding form factors; the branching 
fractions of $\psipto\omega\pi, \rho\eta$ and $\rho\etap$  are listed in
Table \ref{branching}~\cite{emprocess}.

\begin{figure}[hbt]
\centerline{\hbox{\psfig{file=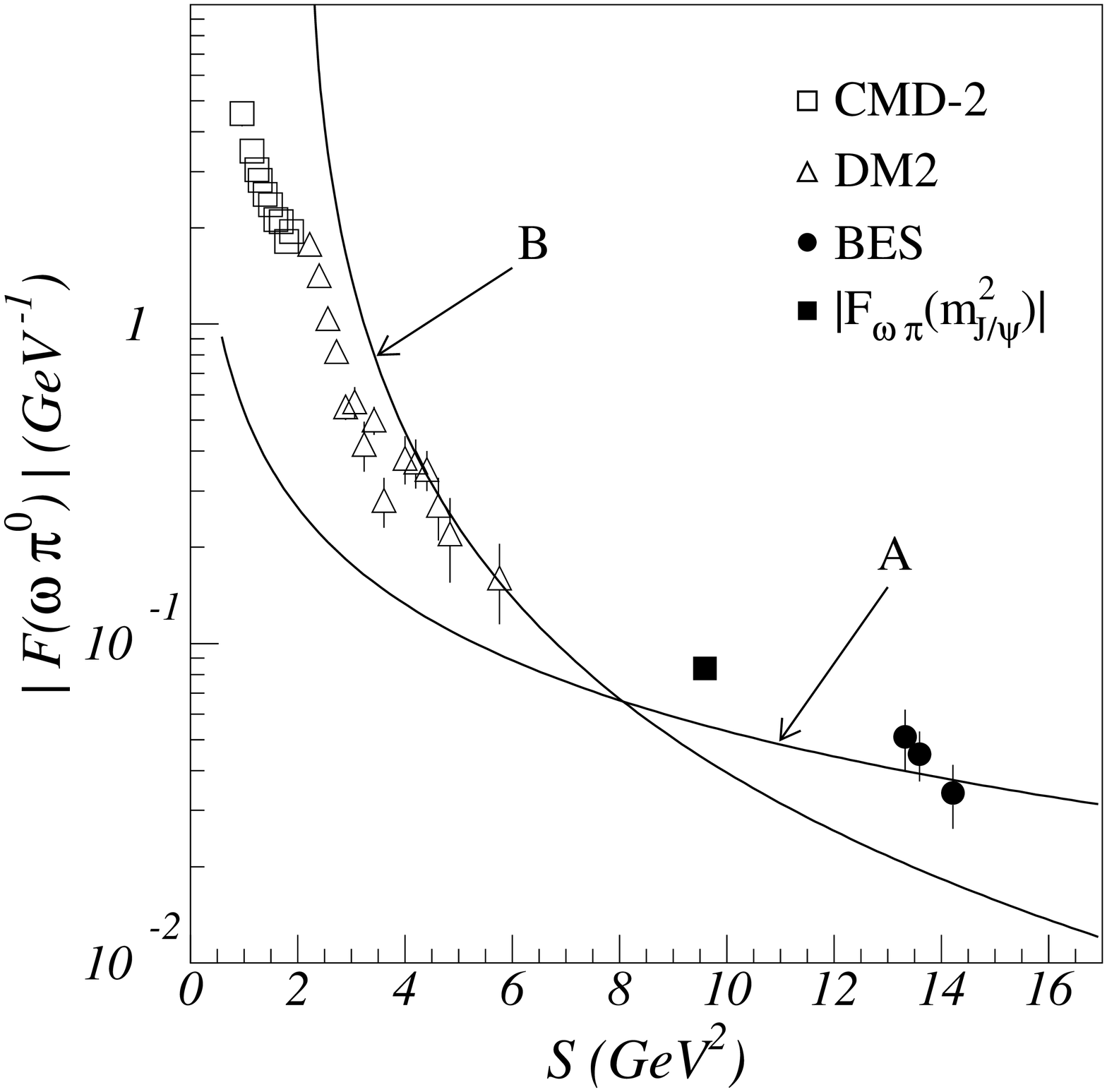,height=6.0cm, width=8.0cm}}}
\caption{\label{Ffwpi} Energy dependence of the $\eeto\gamma^*\to\wpi$ 
form factor. Curve (A) is calculated with Eq. (\ref{Formula2}), 
while curve (B) is calculated with Eq. (\ref{Formula1}). }
\end{figure}

\begin{table}[htbp]
\caption{\label{formfactor} Cross sections and form factors measured for
$\eeto\wpi$, $\rho\eta$, and $\rho\etap$ at $\sqrt{s}=3.650$, 3.686, 
and 3.773 GeV. }
\begin{tabular}{|l|c|c|c|c|c|c|c|} \hline 
Channel & Samples  & $\mathcal{L}$~(pb$^{-1}$)  & $N^{obs}_{Cont.}$ 
& $\epsilon$ (\%) & $1+\delta$ & $\sigma_{0}$ (pb) & 
                           $|\mathcal{F}_{VP}|(\GeV^{-1})$ \\ \hline 
       & 3.650 GeV    & 6.42    & $7.3^{+3.3}_{-2.7}$     &  5.09  & 1.032   
          &  $24.3^{+11.0}_{-9.0}\pm4.3$    &  $0.051^{+0.12}_{-0.10}$   \\  
$\wpi$ & 3.686 GeV    & 19.72   & $17.3^{+5.7}_{-5.1}$    &  4.98  & 1.031   
          & $19.2^{+6.3}_{-5.7}\pm2.9$     &  $0.045^{+0.008}_{-0.007}$   \\  
       & 3.773 GeV    & 17.3    & $8.6^{+4.0}_{-3.3}$  &  5.09     & 1.028
       & $10.7^{+5.0}_{-4.1}\pm1.7$     &  $0.034^{+0.008}_{-0.007}$ \\ \hline
          & 3.650 GeV    & 6.42    & $2.3^{+2.1}_{-1.4}$    &  10.9  & 1.028   
          &  $8.1^{+7.4}_{-4.9}\pm1.1$    &  $0.030^{+0.014}_{-0.009}$   \\  
$\rho\eta$ & 3.686 GeV    & 19.72   & $16.0^{+5.6}_{-5.0}$   &  10.9  & 1.028  
           & $18.4^{+8.6}_{-7.8}\pm1.9$     &  $0.046^{+0.011}_{-0.010}$   \\  
          & 3.773 GeV    & 17.3    & $5.8^{+3.3}_{-2.6}$    &  10.7  & 1.026
          & $7.8^{+4.4}_{-3.5}\pm0.08$ &  $0.030^{+0.009}_{-0.007}$ \\ \hline
           & 3.650 GeV    & 6.42    & $<4.4$    &  4.33  & 1.021   
               &  $<89$    &  $<0.192$     \\  
$\rho\etap$ & 3.686 GeV    & 19.72   & $2.9^{+2.4}_{-1.6}$  &  4.43  & 1.020   
           & $18.6^{+15.4}_{-10.3}\pm3.6$  &  $0.050^{+0.021}_{-0.015}$   \\  
           & 3.773 GeV    & 17.3    & $<3.9$     &  4.56  & 1.019
              & $<28$   &  $<0.106$    \\  \hline 
\end{tabular} \\
\end{table}
Fig. \ref{Ffwpi} shows the results of the form factor 
$|\mathcal{F}_{\wpi}|$
from our measurements, CMD-2~\cite{cmd2},  and DM2~\cite{dm2}, and 
the calculated value of $|\mathcal{F}_{\wpi}|$ 
at $s=m^2_{\jpsi}$.
Curve (A) is  predicted by
J.-M. G\'{e}rard and G. L\'{o}pez Castro~\cite{Gerard} as:
\begin{equation}
|\mathcal{F}_{\omega\pi^0}(s\to \infty) |
=\frac{f_{\omega} f_{\pi}}{3\sqrt{2}s},
\label{Formula2}
\end{equation}
with  $f_{\omega}=17.05\pm0.28$ and $f_{\pi}=0.1307 \GeV$, 
the decay constants of  $\omega$ and $\pi$, respectively. Curve (B) is 
predicted by Victor Chernyak~\cite{chernyak}:
\begin{equation}
|\mathcal{F}_{\omega\pi^0}(s)|  =  |\mathcal{F}_{\omega\pi^0}(0)|
\frac{m^2_{\rho} M^2_{\rho'}}{(m^2_{\rho}- s)(M^2_{\rho'} - s)},
\label{Formula1}
\end{equation}
where $m_{\rho}$ and $M_{\rho'}$ are the masses of $\rho(770)$ 
and $\rho(1450)$, respectively. 
From  Fig. \ref{Ffwpi}, our results agree with
the description of Eq. (\ref{Formula2}).

\section{Measurements of $\psip$ decays into $\kstark$, $\phi \pi^0$, 
$\phi \eta$, $\phi \eta^{'}$, $\omega\eta$, and $\omega \eta^{'}$}
For $\psipto\kstark$, we study its final state 
$\ks K^\pm\pi^\mp\rar\pi^+\pi^-K^\pm\pi^\mp$~\cite{kstark}. The
other decay modes are studied with $\phi$ decays to $K^+K^-$,
$\omega$ to $\pi^+\pi^-\pi^0$, $\etap$ to $\eta\pi^+\pi^-$ or
$\gamma\pi^+\pi^-$, and $\pi^0$ and $\eta$ to 2$\gamma$~\cite{VP}. The
results are listed  in Table \ref{branching}.

\begin{table}[hbt] 
\begin{center}
\caption{ \label{branching} Branching fractions  and upper limits (90\% C.L.) 
measured for $\psip$ decays. Results for corresponding
$\jpsi$ branching fractions and the
ratios $Q_h=\frac{B(\psip\rar h)}{B(\jpsi\rar h)}$ are also given. }
\begin{tabular}{|c|c|c|c|}  \hline  
$h$  & $B(\psi(2S)\rar )\times 10^{-5}$ & $B(J/\psi\rar)\times 10^{-4}$ 
          & $Q_h$ (\%)  \\ \hline
$\rho\pi$ & $5.1\pm0.7\pm1.1$ & 127$\pm$9 & $0.40\pm0.11$  \\ 
$\kstarkpm$  & $2.9^{+1.3}_{-1.7}\pm0.4$ & 50$\pm$4 & $0.59^{+0.27}_{-0.36}$\\
$\kstarknn$  & $13.3^{+2.4}_{-2.8}\pm1.7$ & 42$\pm$4 & $3.2\pm0.8$   \\ \hline
$\wpi$  & $1.87^{+0.68}_{-0.62}\pm0.28$   & 4.2$\pm$0.6   
        & $4.4^{+1.8}_{-1.6}$ \\ 
$\rho\eta$ & $1.78^{+0.67}_{-0.62}\pm0.17$ & 1.93$\pm$0.23 
           & $9.2^{+3.6}_{-3.3}$ \\ 
$\rho\etap$ & $1.87^{+1.64}_{-1.11}\pm0.33$  & 1.05$\pm$0.18 
            & $17.8^{+15.9}_{-11.1}$ \\ \hline
$\phi\pi^0$    & $<0.41$ & $<0.068$ & -- \\ 
$\phi\eta$     & $3.3\pm 1.1\pm0.5$ & 6.5$\pm$0.7 & $5.1\pm 1.9$ \\
$\phi\etap$  & $2.8\pm 1.5\pm0.6$  & 3.3$\pm$0.4 & $8.5 \pm 5.0$ \\
$\omega \eta$   & $<3.2$  & 15.8$\pm$1.6 & $<2.0$  \\
$\omega\etap$  & $3.1^{+2.4}_{-2.0}\pm0.7$ & 1.67$\pm$0.25
               & $19^{+15}_{-13}$ \\  \hline 
$p\bar{p}\pi^0$ & $13.2\pm1.0\pm1.5$ & 10.9$\pm$0.09 & $12.1\pm1.9$ \\
$p\bar{p}\eta$  & $5.8\pm1.1\pm0.7$  & 2.09$\pm$0.18 
                & $ 2.8\pm0.7$ \\ \hline
\end{tabular} 
\end{center}
\end{table}
\section{Analysis of $\psipto p\bar{p}\pi^0(\eta)$}
The final states of these two decay modes are the same $p\bar{p}\gamg$,
and the signal event numbers are got by fitting the $\gamg$ invariant
mass distribution in the selected events with $p\bar{p}\gamg$ final 
state~\cite{ppbar}. The branching fractions for
$\psipto p\bar{p}\pi^0$ and $\psipto p\bar{p}\eta$ are
listed in Table \ref{branching}. For $\psipto p\bar{p}\pi^0$, the errors
are much smaller than the previous measurement by Mark-II~\cite{markII}.
There are enhancements with $p\pi$ and $p\eta$ mass around 1.5 GeV, and 
weak evidences for the $p\bar{p}$ threshold enhancements in both channels.

\section{Summary}
We report the results on $\psip$ decays into 10 VP channels and
 $p\bar{p}\pi^0(\eta)$ final states. The branching fractions 
in our measurement are consistent with those of CLEO~\cite{cleo}.
With the measured branching fractions,
the ``12\% rule'' is tested. From the ratios $Q_h$ in Table \ref{branching},
we see the channels of $\rho\eta$, $\rho\etap$, $\phi\etap$, $\omega\etap$
and $p\bar{p}\pi^0$ are consistent with ``12\% rule'', while the others are
suppressed. The solution to  the ''$\rho\pi$ puzzle'' seems to need
more accurate measurements and  further effort from theory.

\section*{Acknowledgments}
I would like to thank Profs. J. Tran Thanh Van for 
the friendly hospitality. I would like also to thank my colleagues of BES 
collaboration who did the good work which are reported here.

\end{document}